\documentclass[preprint,superscriptaddress,nofootinbib]{revtex4-1}
  \usepackage{amssymb,amsmath,graphicx,subfigure,color}
  \usepackage[colorlinks]{hyperref}
  
  \usepackage{lineno}
  \allowdisplaybreaks[4]
  \begin{document}
  \title{The study of ${\eta}_{c}(1S)$ ${\to}$ $PP^{\prime}$ decays}
  \author{Yueling Yang}
  \affiliation{Institute of Particle and Nuclear Physics,
              Henan Normal University, Xinxiang 453007, China}
  \author{Xule Zhao}
  \affiliation{Institute of Particle and Nuclear Physics,
              Henan Normal University, Xinxiang 453007, China}
  \author{Shuangshi Fang}
  \affiliation{Institute of High Energy Physics,
               Chinese Academy of Sciences, Beijing 100049, China}
  \author{Jinshu Huang}
  \affiliation{School of Physics and Electronic Engineering,
              Nanyang Normal University, Nanyang 473061, China}
  \author{Junfeng Sun}
  \affiliation{Institute of Particle and Nuclear Physics,
              Henan Normal University, Xinxiang 453007, China}

  \begin{abstract}
  The ${\eta}_{c}(1S)$ ${\to}$ $PP^{\prime}$ decays are the
  parity violation modes. These decays can be induced by
  the weak interactions within the standard model, and have
  been searched for based on the available experimental data.
  To meet the needs of experimental investigation,
  the ${\eta}_{c}(1S)$ ${\to}$ $PP^{\prime}$ decays are studied
  with the perturbative QCD approach.
  It is found that branching ratios are the
  order of $10^{-15}$ and less, which offers a ready
  reference for future analyses.
  \end{abstract}
  \maketitle

  Charmonium is a system containing the charmed quark and
  antiquark $c\bar{c}$.
  Recently, the study of charmonium regained a great renewed
  interest due to many new discoveries from the massive
  dedicated investigation by BES-II, CLEO-c, BES-III,
  BaBar, Belle, Belle-II and LHCb \cite{pdg2020}.

  The ${\eta}_{c}(1S)$ meson is commonly referred to as
  ${\eta}_{c}$.
  Both the total spin and orbital angular momentum
  of $c$ and $\bar{c}$ quarks in ${\eta}_{c}$ are zero.
  The ${\eta}_{c}$ particle is the paracharmonium state with
  the well established quantum number of $J^{PC}$ $=$ $0^{-+}$
  \cite{pdg2020}.
  Its $J^{PC}$ is different from that of photon.
  ${\eta}_{c}$ cannot be directly produced at the $e^{+}e^{-}$
  collisions.
  However, ${\eta}_{c}$ can be produced via the magnetic dipole
  transition process $J/{\psi}$ ${\to}$ ${\gamma}{\eta}_{c}$
  with the branching ratio of
  ${\cal B}r(J/{\psi}{\to}{\gamma}{\eta}_{c})$ $=$ $(1.7{\pm}0.4)\%$
  \cite{pdg2020}.
  Up to now, there is over $10^{10}$ $J/{\psi}$ data samples
  with BESIII detector \cite{dataweb}, the largest amount of
  available statistics, and corresponding to more than
  $10^{8}$ ${\eta}_{c}$.
  Given the large $J/{\psi}$ production cross section ${\sigma}$ ${\sim}$
  $3400$ $nb$ \cite{nimpra614.345}, it is expected that more than
  $10^{13}$ $J/{\psi}$, corresponding to more than $10^{11}$
  ${\eta}_{c}$, could be accumulated at the Super Tau Charm
  Facility (STCF) with $3\,ab^{-1}$ on-resonance dataset
  in the future.
  This provides a good opportunity for studying the properties
  of ${\eta}_{c}$ particle.

  Although there is a large amount of data, the experimental
  study of ${\eta}_{c}$ decays is comparatively limited.
  So far, only 33 exclusive ${\eta}_{c}$ decay modes have
  been reported with concrete numerical value.
  The sum of the 33 branching ratios is about 63\%, and most
  of measurements have very large uncertainties \cite{pdg2020}.
  The mass of ${\eta}_{c}$ particle, $m_{{\eta}_{c}}$ $=$
  $2983.9{\pm}0.5$ MeV \cite{pdg2020}, is minimal among
  charmonium, and lies below the open charm threshold.
  So the ${\eta}_{c}$ decay into hadronic states through the
  strong interactions is severely hindered by the phenomenological
  Okubo-Zweig-Iizuka (OZI) rule \cite{ozi-o,ozi-z,ozi-i}.
  The $c\bar{c}$ quark pair in the ${\eta}_{c}$ state
  can annihilate into two gluons and two photons with branching
  ratio of ${\cal B}r({\eta}_{c}{\to}{\gamma}{\gamma})$ $=$
  $(1.58{\pm}0.11){\times}10^{-4}$ \cite{pdg2020}.
  Among the nonleptonic ${\eta}_{c}$ decays, the simplest
  hadronic final states are two pseudoscalar mesons.
  However, it should be pointed out that the ${\eta}_{c}$
  ${\to}$ $PP^{\prime}$ decays are the parity violating modes,
  so they should be induced by  the weak interactions rather
  than the strong and electromagnetic ones.
  The ${\eta}_{c}$ ${\to}$ $PP^{\prime}$ decays were
  experimentally studied at BES-II and BES-III, but no significant
  signals are observed and only the upper limits on
  branching ratios are obtained by now
  \cite{pdg2020,prd84.032006,epjc45.337}.
  As far as we know, there is no theoretical investigation
  on the ${\eta}_{c}$ ${\to}$ $PP^{\prime}$ decays yet.
  In this paper, according to the future experimental prospects,
  we will study the ${\eta}_{c}$ ${\to}$ $PP^{\prime}$ decays
  within the standard model (SM) of the elementary particles
  in order to offer a ready reference for future analysis.

  Within SM, the ${\eta}_{c}$ ${\to}$ $PP^{\prime}$ decays
  are induced by the $W^{\pm}$ exchange interaction.
  At the quark level, based on the operator product expansion
  and renormalization group (RG) method, the effective
  Hamiltonian in charge of the ${\eta}_{c}$ ${\to}$
  $PP^{\prime}$ decays is written as \cite{rmp68.1125},
   \begin{equation}
  {\cal H}_{\rm eff}\ =\
   \frac{G_{F}}{\sqrt{2}}\, \sum\limits_{q_{1},q_{2}}\,
   V_{cq_{1}}\,V_{cq_{2}}^{\ast}\,
   \big\{ C_{1}({\mu})\,O_{1}({\mu})
         +C_{2}({\mu})\,O_{2}({\mu}) \big\}
         +{\rm h.c.}
   \label{eq:hamilton},
   \end{equation}
  where $G_{F}$ ${\simeq}$ $1.166{\times}10^{-5}\,{\rm GeV}^{-2}$
  \cite{pdg2020} is the Fermi coupling constant, and
  $q_{1,2}$ $=$ $d$ and $s$.
  The averaged values of the Cabibbo-Kobayashi-Maskawa (CKM)
  elements are ${\vert}V_{cs}{\vert}$ $=$ $0.987(11)$ and
  ${\vert}V_{cd}{\vert}$ $=$ $0.221(4)$ \cite{pdg2020}.
  The parameter ${\mu}$ is a factorization scale, which
  divides the physical contributions into two parts,
  the short- and long-distance contributions.
  The Wilson coefficients $C_{1,2}$ summarize the short-distance
  physical contributions above the scales of ${\mu}$.
  They are computable with the RG-improved perturbation theory
  at the scale of the mass of gauge $W$ boson, $m_{W}$, and
  then evolved to a characteristic scale of ${\mu}$ for $c$
  quark decay.
  \begin{equation}
  \vec{C}({\mu})\, =\,
  U_{4}({\mu},m_{b})\,U_{5}(m_{b},m_{W})\, \vec{C}(m_{W})
  \label{ci},
  \end{equation}
  where the explicit expression of $U_{f}({\mu}_{f},{\mu}_{i})$
  can be found in Ref. \cite{rmp68.1125}.
  The operators describing the local interactions among four
  quarks are defined as,
   \begin{eqnarray}
   O_{1} &=&
    \big[ \bar{c}_{\alpha}\,{\gamma}_{\mu}\,
         (1-{\gamma}_{5})\,q_{1,{\alpha}} \big]\,
    \big[ \bar{q}_{2,{\beta}}\,{\gamma}^{\mu}\,
         (1-{\gamma}_{5})\,c_{\beta} \big]
   \label{operator-01}, \\
   O_{2} &=&
    \big[ \bar{c}_{\alpha}\,{\gamma}_{\mu}\,
         (1-{\gamma}_{5})\,q_{1,{\beta}} \big]\,
    \big[ \bar{q}_{2,{\beta}}\,{\gamma}^{\mu}\,
         (1-{\gamma}_{5})\,c_{\alpha} \big]
   \label{operator-02},
   \end{eqnarray}
  where ${\alpha}$ and ${\beta}$ are color indices.
  The contributions of penguin operators are neglected because
  of the strong suppression from the CKM factors,
  $(V_{uq_{1}}\,V_{uq_{2}}^{\ast}+V_{cq_{1}}\,V_{cq_{2}}^{\ast})/
   (V_{cq_{1}}\,V_{cq_{2}}^{\ast})$ $=$
  $-(V_{tq_{1}}\,V_{tq_{2}}^{\ast})/
    (V_{cq_{1}}\,V_{cq_{2}}^{\ast})$ $=$ ${\cal O}({\lambda}^{4})$
  with the Wolfenstein parameter ${\lambda}$ ${\approx}$ $0.2$.

  The decay amplitudes can be written as,
   \begin{equation}
  {\cal A}({\eta}_{c}{\to}PP^{\prime})\, =\,
  {\langle}PP^{\prime}{\vert} {\cal H}_{\rm eff}
  {\vert}{\eta}_{c} {\rangle}\, =\,
   \frac{G_{F}}{\sqrt{2}}\, \sum\limits_{q_{1},q_{2}}\,
   V_{cq_{1}}\,V_{cq_{2}}^{\ast}\, \sum\limits_{i=1}^{2}\,
   C_{i}({\mu})\,{\langle}PP^{\prime}{\vert} O_{i}({\mu})\,
  {\vert} {\eta}_{c} {\rangle}
   \label{eq:hamilton-amplitudes}.
   \end{equation}
  In Eq.(\ref{eq:hamilton-amplitudes}), the Fermi constant $G_{F}$,
  the CKM elements ${\vert}V_{cs}{\vert}$ and ${\vert}V_{cd}{\vert}$
  have been pretty well determined from data, and the Wilson
  coefficients $C_{1,2}$ could be reliably computed.
  So the remaining theoretical work is the evaluations of hadronic
  matrix elements (HMEs) ${\langle}O_{i}{\rangle}$ $=$
  ${\langle}PP^{\prime}{\vert} O_{i}({\mu})\,
  {\vert} {\eta}_{c} {\rangle}$.
  HMEs describe the complex transformation between quarks and
  hadrons, and contain the perturbative and nonperturbative
  contributions.

  Recently, some QCD-inspired phenomenological models, such as
  the QCD factorization (QCDF) approach \cite{prl83.1914,
  npb591.313,npb606.245,plb488.46,plb509.263,prd64.014036}
  and the perturbative QCD (pQCD) approach
  \cite{prl74.4388,plb348.597,prd52.3958,prd63.074006,
  prd63.054008,prd63.074009,plb555.197}, have been
  technically proposed and widely applied to evaluate HMEs.
  With these phenomenological models, HMEs are generally
  written as the convolution of scattering amplitudes
  and the hadronic wave functions (WFs).
  The scattering amplitudes and WFs reflect the contributions
  at the quark and hadron levels, respectively.
  The scattering amplitudes arising from hard gluon exchanges
  among quarks are calculable with the perturbative theory.
  WFs including the momentum distributions of hadronic
  compositions are universal, and could be obtained by
  nonperturbative methods or from data.
  In the practical theoretical calculation, the transverse
  momentum and Sudakov factors are proposed by the pQCD approach
  to provide an effective cutoff for the endpoint singularities
  from the collinear approximation.
  In this paper, we will investigate the ${\eta}_{c}$ ${\to}$
  $PP^{\prime}$ decays with the pQCD approach, where
  the decay amplitudes are expressed as the convolution integral
  of three parts : the Wilson coefficients $C_{i}$, scattering
  amplitudes ${\cal H}$ and hadronic WFs ${\Phi}$.
   \begin{eqnarray}
  {\cal A}_{i} &=& {\int} dx_{1}\,dx_{2}\,dx_{3}\,
   db_{1}\,db_{2}\,db_{3}\,C_{i}(t_{i})\,
  {\cal H}_{i}(x_{1},x_{2},x_{3},b_{1},b_{2},b_{3})
   \nonumber \\ & & \quad
  {\Phi}_{{\eta}_{c}}(x_{1},b_{1})\,e^{-S_{{\eta}_{c}}}\,
  {\Phi}_{P}(x_{2},b_{2})\,e^{-S_{P}}\,
  {\Phi}_{P^{\prime}}(x_{3},b_{3})\,e^{-S_{P^{\prime}}}
   \label{eq:pqcd-hme},
   \end{eqnarray}
  where $x_{i}$ is the longitudinal momentum fraction of the valence
  quark, $b_{i}$ is the conjugate variable of the transverse momentum,
  and $e^{-S_{i}}$ is the Sudakov factor.

\begin{figure}[ht]
  \includegraphics[width=0.22\textwidth,bb=200 535 370 635]{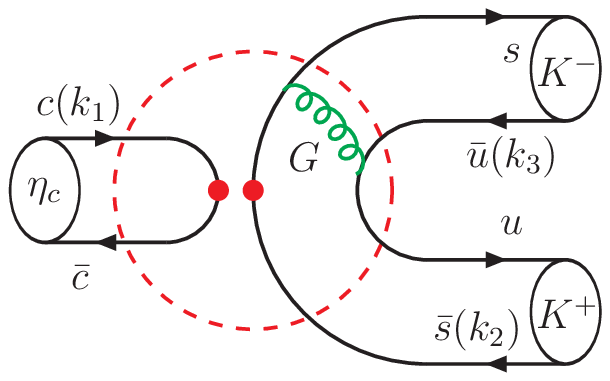}\quad
  \includegraphics[width=0.22\textwidth,bb=200 535 370 635]{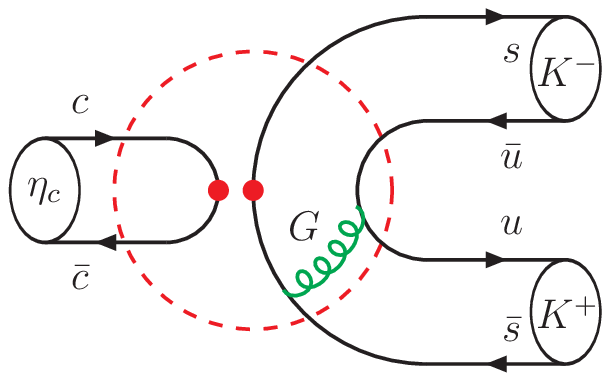}\quad
  \includegraphics[width=0.22\textwidth,bb=200 535 370 635]{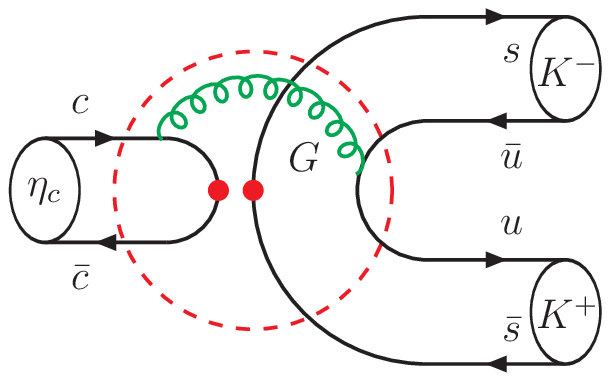}\quad
  \includegraphics[width=0.22\textwidth,bb=200 535 370 635]{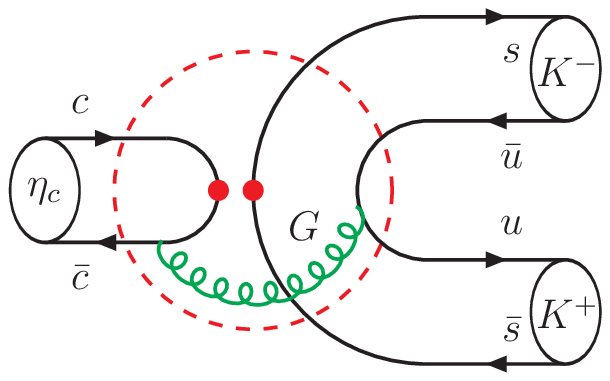} \\
  {(a) \hspace{0.21\textwidth} (b)
       \hspace{0.21\textwidth} (c)
       \hspace{0.21\textwidth} (d)}
  \caption{The Feynman diagrams for the ${\eta}_{c}$ ${\to}$
  $K^{-}K^{+}$ decay with the pQCD approach, where (a,b) are
  factorizable diagrams, and (c,d) are nonfactorizable diagrams.
  The dots denote appropriate interactions, and the dashed circles
  denote scattering amplitudes.}
  \label{feynman-pqcd}
  \end{figure}

  In the calculation, it is convenient to use the light-cone
  vectors to define the kinematic variables.
  In the rest frame of the ${\eta}_{c}$ meson, one has
   \begin{equation}
    p_{{\eta}_{c}}\, =\, p_{1}\, =\,
   \frac{m_{{\eta}_{c}}}{\sqrt{2}}(1,1,0)
   \label{kine-etac},
   \end{equation}
   \begin{equation}
    p_{P}\, =\, p_{2}\, =\,
   \frac{m_{{\eta}_{c}}}{\sqrt{2}}(1,0,0)
   \label{kine-p1},
   \end{equation}
   \begin{equation}
   p_{P^{\prime}}\, =\, p_{3}\, =\,
   \frac{m_{{\eta}_{c}}}{\sqrt{2}}(0,1,0)
   \label{kine-p2},
   \end{equation}
   \begin{equation}
    k_{1}\, =\,
   \frac{m_{{\eta}_{c}}}{\sqrt{2}}(x_{1},x_{1},\vec{k}_{1T})
   \label{kine-k1},
   \end{equation}
   \begin{equation}
   k_{2}\, =\,
   \frac{m_{{\eta}_{c}}}{\sqrt{2}}(x_{2},0,\vec{k}_{2T})
   \label{kine-k2},
   \end{equation}
   \begin{equation}
   k_{3}\, =\,
   \frac{m_{{\eta}_{c}}}{\sqrt{2}}(0,x_{3},\vec{k}_{3T})
   \label{kine-k3},
   \end{equation}
   where $k_{i}$, $x_{i}$ and $\vec{k}_{iT}$ are respectively
   the momentum, longitudinal momentum fraction and
   transverse momentum, as shown in Fig. \ref{feynman-pqcd} (a).

  With the convention of Refs. \cite{epjc73.2437,jhep0605.004,
  prd65.014007,2012.10581}, the WFs and distribution amplitudes
  (DAs) are defined as follows.
   \begin{equation}
  {\langle}\,0\,{\vert}\,\bar{c}_{\alpha}(0)c_{\beta}(z)\,
  {\vert} {\eta}_{c}(p_{1})\,{\rangle}\, =\,
  -\frac{i}{4}\,f_{{\eta}_{c}}\,
  {\int}_{0}^{1}\,dx_{1}\, e^{-i\,k_{1}{\cdot}z}\, \big\{
   \big[\!\not{p}_{1}\, {\phi}_{{\eta}_{c}}^{a}
   +m_{{\eta}_{c}}\, {\phi}_{{\eta}_{c}}^{p} \big]\,
  {\gamma}_{5} \big\}_{{\beta}{\alpha}}
   \label{wave-cc},
   \end{equation}
    \begin{eqnarray} & &
   {\langle}\,P(p_{2})\,{\vert}\, \bar{q}_{\alpha}(0)\,
   q_{1{\beta}}(z)\, {\vert}\,0\,{\rangle}
    \nonumber \\ &=&
   -\frac{i\,f_{P}}{4}\,
   {\int}_{0}^{1}\,dx_{1}\, e^{+i\,k_{2}{\cdot}z}\,
    \big\{ {\gamma}_{5}\, \big[ \!\not{p}_{2}\,
   {\phi}_{P}^{a} +{\mu}_{P}\,{\phi}_{P}^{p}
   -{\mu}_{P}\,\big( \!\not{n}_{+}\!\not{n}_{-}-1\big)\,
    {\phi}_{P}^{t} \big] \big\}_{{\beta}{\alpha}}
    \label{wf-pseudoscalar-01},
    \end{eqnarray}
    \begin{eqnarray} & &
   {\langle}\,P^{\prime}(p_{3})\,{\vert}\,
    \bar{q}_{2{\alpha}}(0)\, q_{\beta}(z)\, {\vert}\,0\,{\rangle}
    \nonumber \\ &=&
   -\frac{i\,f_{P^{\prime}}}{4}\,
   {\int}_{0}^{1}\,dx_{1}\, e^{+i\,k_{3}{\cdot}z}\,
    \big\{ {\gamma}_{5}\, \big[ \!\not{p}_{3}\,
   {\phi}_{P^{\prime}}^{a}
   +{\mu}_{P^{\prime}}\,{\phi}_{P^{\prime}}^{p}
   -{\mu}_{P^{\prime}}\,\big( \!\not{n}_{-}\!\not{n}_{+}-1\big)\,
    {\phi}_{P^{\prime}}^{t} \big] \big\}_{{\beta}{\alpha}}
    \label{wf-pseudoscalar-02},
    \end{eqnarray}
  where $f_{{\eta}_{c}}$ and $f_{P,P^{\prime}}$ are decay constants.
  ${\mu}_{P,P^{\prime}}$ $=$ $1.6{\pm}0.2$ GeV \cite{jhep0605.004}
  is the chiral mass.
  $n_{+}$ $=$ $(1,0,0)$ and $n_{-}$ $=$ $(0,1,0)$ are the
  null vectors.
  The explicit expressions of ${\phi}_{{\eta}_{c}}^{a,p}$
  and ${\phi}_{P}^{a,p,t}$ can be found in Ref. \cite{epjc73.2437}
  and Refs. \cite{2012.10581,jhep0605.004}, respectively.
  We collect and display these WFs and DAs as follows.
    \begin{equation}
   {\phi}_{{\eta}_{c}}^{a}(x,b)\, =\,
    N^{a}\,x\,\bar{x}\,{\exp} \Big\{ -\frac{m_{c}}{\omega}\,
    x\,\bar{x}\, \Big[ \Big(
    \frac{x-\bar{x}}{2\,x\,\bar{x}} \Big)^{2}+
   {\omega}^{2}\,b^{2} \Big] \Big\}
    \label{wf-etac-a},
    \end{equation}
    \begin{equation}
   {\phi}_{{\eta}_{c}}^{p}(x,b)\, =\,
    N^{p}\, {\exp} \Big\{ -\frac{m_{c}}{\omega}\,x\,\bar{x}\,
    \Big[ \Big(\frac{x-\bar{x}}{2\,x\,\bar{x}}\Big)^{2}
  +{\omega}^{2}\,b^{2} \Big] \Big\}
    \label{wf-etac-p},
    \end{equation}
    \begin{equation}
   {\phi}_{P}^{a}(x)\, =\, 6\,x\,\bar{x}\,\big\{
    1+a_{1}^{P}\,C_{1}^{3/2}({\xi})
     +a_{2}^{P}\,C_{2}^{3/2}({\xi})\big\}
    \label{wf-pi-twsit-2},
    \end{equation}
    \begin{eqnarray}
   {\phi}_{P}^{p}(x) &=& 1+3\,{\rho}_{+}^{P}
   -9\,{\rho}_{-}^{P}\,a_{1}^{P}
   +18\,{\rho}_{+}^{P}\,a_{2}^{P}
    \nonumber \\ &+&
    \frac{3}{2}\,({\rho}_{+}^{P}+{\rho}_{-}^{P})\,
    (1-3\,a_{1}^{P}+6\,a_{2}^{P})\,{\ln}(x)
    \nonumber \\ &+&
    \frac{3}{2}\,({\rho}_{+}^{P}-{\rho}_{-}^{P})\,
    (1+3\,a_{1}^{P}+6\,a_{2}^{P})\,{\ln}(\bar{x})
    \nonumber \\ &-&
    (\frac{3}{2}\,{\rho}_{-}^{P}
    -\frac{27}{2}\,{\rho}_{+}^{P}\,a_{1}^{P}
    +27\,{\rho}_{-}^{P}\,a_{2}^{P})\,C_{1}^{1/2}(\xi)
    \nonumber \\ &+&
    ( 30\,{\eta}_{P}-3\,{\rho}_{-}^{P}\,a_{1}^{P}
    +15\,{\rho}_{+}^{P}\,a_{2}^{P})\,C_{2}^{1/2}(\xi)
    \label{wf-pi-twsit-3-p},
    \end{eqnarray}
    \begin{eqnarray}
   {\phi}_{P}^{t}(x) &=&
    \frac{3}{2}\,({\rho}_{-}^{P}-3\,{\rho}_{+}^{P}\,a_{1}^{P}
    +6\,{\rho}_{-}^{P}\,a_{2}^{P})
    \nonumber \\ &-&
    C_{1}^{1/2}(\xi)\big\{
    1+3\,{\rho}_{+}^{P}-12\,{\rho}_{-}^{P}\,a_{1}^{P}
   +24\,{\rho}_{+}^{P}\,a_{2}^{P}
    \nonumber \\ & & \quad +
    \frac{3}{2}\,({\rho}_{+}^{P}+{\rho}_{-}^{P})\,
    (1-3\,a_{1}^{P}+6\,a_{2}^{P})\,{\ln}(x)
    \nonumber \\ & & \quad +
    \frac{3}{2}\,({\rho}_{+}^{P}-{\rho}_{-}^{P})\,
    (1+3\,a_{1}^{P}+6\,a_{2}^{P})\, {\ln}(\bar{x}) \big\}
    \nonumber \\ &-&
    3\,(3\,{\rho}_{+}^{P}\,a_{1}^{P}
    -\frac{15}{2}\,{\rho}_{-}^{P}\,a_{2}^{P})\,C_{2}^{1/2}(\xi)
    \label{wf-pi-twsit-3-t},
    \end{eqnarray}
   where $\bar{x}$ $=$ $1$ $-$ $x$ and ${\xi}$ $=$ $x$ $-$
   $\bar{x}$ $=$ $2\,x$ $-$ $1$.
   ${\omega}$ $=$ $m_{c}\,{\alpha}_{s}(m_{c})$ is the shape
   parameter.
   The parameters $N^{a,p}$ is determined by the normalization
   conditions,
   \begin{equation}
  {\int} {\phi}_{{\eta}_{c}}^{a,p}(x,0)\,dx\, =\, 1
   \label{normalization-etac-ap}.
   \end{equation}
   The meaning and definition of other parameters can refer
   to Refs. \cite{2012.10581,jhep0605.004}.

   From Fig. \ref{feynman-pqcd}, it can be clearly seen that
   there are only annihilation amplitudes for the ${\eta}_{c}$
   ${\to}$ $PP^{\prime}$ decays in SM, because the valence
   quarks of the final states are entirely different from
   those of the initial state.
   The annihilation contributions are necessary and important in
   nonleptonic two-body $B$ meson decays \cite{prd65.074001,
   prd65.094025,prd68.054003,npb675.333,npb774.64,prd90.054019,
   prd91.074026,plb740.56,plb743.444,plb504.6,prd70.034009,
   prd85.094003,prd76.074018,prd88.014043}.
   The ${\eta}_{c}$ ${\to}$ $PP^{\prime}$ decays offer another
   processes to investigate the annihilation contributions
   within the factorization approaches, besides the $B_{d}$
   ${\to}$ $K^{+}K^{-}$ and $B_{s}$ ${\to}$ ${\pi}{\pi}$ decays.
   The decay amplitudes are written as follows.
    \begin{equation}
   {\cal A}({\eta}_{c}{\to}K^{+}K^{-})\, =\,
    \frac{G_{F}}{\sqrt{2}}\, V_{cs}\,V_{cs}^{\ast}\,
    \big\{ a_{2}\,{\cal A}_{ab}(\overline{K},K)
          +C_{1}\,{\cal A}_{cd}(\overline{K},K) \big\}
    \label{amp-kp-km},
    \end{equation}
    \begin{eqnarray}
   {\cal A}({\eta}_{c}{\to}K^{0}\overline{K}^{0}) &=&
    \frac{G_{F}}{\sqrt{2}}\, \big\{
    V_{cs}\,V_{cs}^{\ast}\,
    \big[ a_{2}\,{\cal A}_{ab}(\overline{K},K)
          +C_{1}\,{\cal A}_{cd}(\overline{K},K) \big]
    \nonumber \\ & & \hspace{0.04\textwidth}
   +V_{cd}\,V_{cd}^{\ast}\,
    \big[ a_{2}\,{\cal A}_{ab}(K,\overline{K})
         +C_{1}\,{\cal A}_{cd}(K,\overline{K}) \big] \big\}
    \label{amp-kz-kzb},
    \end{eqnarray}
    \begin{equation}
   {\cal A}({\eta}_{c}{\to}{\pi}^{+}K^{-})\, =\,
    \frac{G_{F}}{\sqrt{2}}\, V_{cd}\,V_{cs}^{\ast}\,
    \big\{ a_{2}\,{\cal A}_{ab}(\overline{K},{\pi})
          +C_{1}\,{\cal A}_{cd}(\overline{K},{\pi}) \big\}
    \label{amp-km-pip},
    \end{equation}
    \begin{equation}
   {\cal A}({\eta}_{c}{\to}{\pi}^{0}\overline{K}^{0})\, =\,
   -\frac{G_{F}}{2}\, V_{cd}\,V_{cs}^{\ast}\,
    \big\{ a_{2}\,{\cal A}_{ab}(\overline{K},{\pi})
          +C_{1}\,{\cal A}_{cd}(\overline{K},{\pi}) \big\}
    \label{amp-kzb-piz},
    \end{equation}
    \begin{equation}
   {\cal A}({\eta}_{c}{\to}{\pi}^{+}{\pi}^{-})\, =\,
    \frac{G_{F}}{\sqrt{2}}\, V_{cd}\,V_{cd}^{\ast}\,
    \big\{ a_{2}\,{\cal A}_{ab}({\pi},{\pi})
          +C_{1}\,{\cal A}_{cd}({\pi},{\pi}) \big\}
    \label{amp-pip-pim},
    \end{equation}
    \begin{equation} \sqrt{2}\,
   {\cal A}({\eta}_{c}{\to}{\pi}^{0}{\pi}^{0})\, =\,
    \frac{G_{F}}{\sqrt{2}}\, V_{cd}\,V_{cd}^{\ast}\,
    \big\{ a_{2}\,{\cal A}_{ab}({\pi},{\pi})
          +C_{1}\,{\cal A}_{cd}({\pi},{\pi}) \big\}
    \label{amp-piz-piz},
    \end{equation}
    \begin{equation}
   {\cal A}({\eta}_{c}{\to}\overline{K}^{0}{\eta}_{s})\, =\,
   \frac{G_{F}}{\sqrt{2}}\, V_{cd}\,V_{cs}^{\ast}\,
    \big\{ a_{2}\,{\cal A}_{ab}({\eta}_{s},\overline{K})
    +C_{1}\, {\cal A}_{cd}({\eta}_{s},\overline{K}) \big\}
    \label{amp-kzbar-etas},
    \end{equation}
    \begin{equation}
   {\cal A}({\eta}_{c}{\to}\overline{K}^{0}{\eta}_{q}) \, =\,
   \frac{G_{F}}{2}\, V_{cd}\,V_{cs}^{\ast}\,
    \big\{ a_{2}\,{\cal A}_{ab}(\overline{K},{\eta}_{q})
    +C_{1}\, {\cal A}_{cd}(\overline{K},{\eta}_{q}) \big\}
    \label{amp-kzbar-etaq},
    \end{equation}
    \begin{equation}
   {\cal A}({\eta}_{c}{\to}\overline{K}^{0}{\eta}) \, =\,
   {\cal A}({\eta}_{c}{\to}\overline{K}^{0}{\eta}_{q})\,
   {\cos}{\phi}
  -{\cal A}({\eta}_{c}{\to}\overline{K}^{0}{\eta}_{s})\,
   {\sin}{\phi}
    \label{amp-kzbar-eta},
    \end{equation}
    \begin{equation}
   {\cal A}({\eta}_{c}{\to}\overline{K}^{0}{\eta}^{\prime}) \, =\,
   {\cal A}({\eta}_{c}{\to}\overline{K}^{0}{\eta}_{q})\,
   {\sin}{\phi}
  +{\cal A}({\eta}_{c}{\to}\overline{K}^{0}{\eta}_{s})\,
   {\cos}{\phi}
    \label{amp-kzbar-eta-prime},
    \end{equation}
    \begin{eqnarray}
   {\cal A}({\eta}_{c}{\to}{\pi}^{0}{\eta}_{q}) &=&
   -\frac{1}{2}\,\frac{G_{F}}{\sqrt{2}}\, V_{cd}\,V_{cd}^{\ast}\,
    \big\{ a_{2}\,\big[ {\cal A}_{ab}({\pi},{\eta}_{q})
    +{\cal A}_{ab}({\eta}_{q},{\pi}) \big]
    \nonumber \\ & & \qquad \qquad \qquad
    +C_{1}\, \big[  {\cal A}_{cd}({\pi},{\eta}_{q})
    +{\cal A}_{cd}({\eta}_{q},{\pi}) \big]  \big\}
    \label{amp-piz-etaq},
    \end{eqnarray}
    \begin{equation}
   {\cal A}({\eta}_{c}{\to}{\pi}^{0}{\eta}) \, =\,
   {\cal A}({\eta}_{c}{\to}{\pi}^{0}{\eta}_{q})\,{\cos}{\phi}
    \label{amp-piz-eta},
    \end{equation}
    \begin{equation}
   {\cal A}({\eta}_{c}{\to}{\pi}^{0}{\eta}^{\prime}) \, =\,
   {\cal A}({\eta}_{c}{\to}{\pi}^{0}{\eta}_{q})\,{\sin}{\phi}
    \label{amp-piz-eta-prime},
    \end{equation}
    \begin{equation}
   {\cal A}({\eta}_{c}{\to}{\eta}_{s}{\eta}_{s})\, =\,
    \sqrt{2}\,G_{F}\, V_{cs}\,V_{cs}^{\ast}\,
    \big\{ a_{2}\,{\cal A}_{ab}({\eta}_{s},{\eta}_{s})
          +C_{1}\,{\cal A}_{cd}({\eta}_{s},{\eta}_{s}) \big\}
    \label{amp-etas-etas},
    \end{equation}
    \begin{equation}
   {\cal A}({\eta}_{c}{\to}{\eta}_{q}{\eta}_{q})\, =\,
    \frac{G_{F}}{\sqrt{2}}\, V_{cd}\,V_{cd}^{\ast}\,
    \big\{ a_{2}\,{\cal A}_{ab}({\eta}_{q},{\eta}_{q})
          +C_{1}\,{\cal A}_{cd}({\eta}_{q},{\eta}_{q}) \big\}
    \label{amp-etaq-etaq},
    \end{equation}
    \begin{equation}  \sqrt{2}\,
   {\cal A}({\eta}_{c}{\to}{\eta}{\eta})\, =\,
   {\cal A}({\eta}_{c}{\to}{\eta}_{q}{\eta}_{q})\,{\cos}^{2}{\phi}
  +{\cal A}({\eta}_{c}{\to}{\eta}_{s}{\eta}_{s})\,{\sin}^{2}{\phi}
    \label{amp-eta-eta},
    \end{equation}
    \begin{equation}
   {\cal A}({\eta}_{c}{\to}{\eta}{\eta}^{\prime})\, =\,
    \big\{ {\cal A}({\eta}_{c}{\to}{\eta}_{q}{\eta}_{q})
  -{\cal A}({\eta}_{c}{\to}{\eta}_{s}{\eta}_{s}) \big\}\,
   {\sin}{\phi}\,{\cos}{\phi}
    \label{amp-eta-eta-prime},
    \end{equation}
    \begin{equation} \sqrt{2}\,
   {\cal A}({\eta}_{c}{\to}{\eta}^{\prime}{\eta}^{\prime})\, =\,
   {\cal A}({\eta}_{c}{\to}{\eta}_{q}{\eta}_{q})\,{\sin}^{2}{\phi}
  +{\cal A}({\eta}_{c}{\to}{\eta}_{s}{\eta}_{s})\,{\cos}^{2}{\phi}
    \label{amp-eta-prime-eta-prime},
    \end{equation}
  where the amplitude building blocks ${\cal A}_{ij}$ are
  listed in Appendix \ref{blocks}.
  As for the above decay amplitudes, there are two comments.

  (1)
  The isoscalar states ${\eta}$ and/or ${\eta}^{\prime}$ with
  the same $J^{PC}$ are mixtures of the $SU(3)$ octet and
  singlet states. In our calculation, we will adopt the
  quark-flavor basis description proposed in Ref.
  \cite{prd58.114006}, {\em i.e.},
   \begin{equation}
   \left( \begin{array}{c}
   {\eta} \\ {\eta}^{\prime} \end{array} \right)\, =\,
   \left(\begin{array}{cc}
  {\cos}{\phi} & -{\sin}{\phi} \\
  {\sin}{\phi} &  {\cos}{\phi}
   \end{array} \right)\,
   \left( \begin{array}{c}
  {\eta}_{q} \\ {\eta}_{s}
   \end{array} \right)
   \label{amp-eta-mixing-01},
   \end{equation}
  where the mixing angle is ${\phi}$ $=$ $39.3(1.0)^{\circ}$,
  and the flavor bases are
  ${\eta}_{q}$ $=$ $(u\bar{u}+d\bar{d})/{\sqrt{2}}$
  and ${\eta}_{s}$ $=$ $s\bar{s}$ \cite{prd58.114006}.
  Here, it is assumed that (a) the components of glueball,
  charmonium or bottomonium are negligible, and (b) that
  the DAs of ${\eta}_{q}$ and ${\eta}_{s}$ are the same as
  those of ${\pi}$ meson, but with different decay
  constants and mass \cite{prd58.114006,prd76.074018,prd89.114019},
   \begin{equation}
   f_{q}\, =\, 1.07(2)\,f_{\pi}
   \label{decay-constant-etaq},
   \end{equation}
   \begin{equation}
   f_{s}\, =\, 1.34(6)\, f_{\pi}
   \label{decay-constant-etas},
   \end{equation}
   \begin{equation}
   m_{{\eta}_{q}}^{2}\, =\,
   m_{\eta}^{2}\,{\cos}^{2}{\phi}
  +m_{{\eta}^{\prime}}^{2}\,{\sin}^{2}{\phi}
  -\frac{\sqrt{2}\,f_{s}}{f_{q}}
  (m_{{\eta}^{\prime}}^{2}- m_{\eta}^{2})\,
  {\cos}{\phi}\,{\sin}{\phi}
   \label{mass-etaq},
   \end{equation}
   \begin{equation}
   m_{{\eta}_{s}}^{2}\, =\,
   m_{\eta}^{2}\,{\sin}^{2}{\phi}
  +m_{{\eta}^{\prime}}^{2}\,{\cos}^{2}{\phi}
  -\frac{f_{q}}{\sqrt{2}\,f_{s}}
  (m_{{\eta}^{\prime}}^{2}- m_{\eta}^{2})\,
  {\cos}{\phi}\, {\sin}{\phi}
   \label{mass-etas},
   \end{equation}

  (2)
  One distinctive feature is that the amplitudes for
  ${\eta}_{c}$ ${\to}$ $K^{+}K^{-}$ and ${\pi}^{+}{\pi}^{-}$
  decays are respectively proportional to the module square
  of the CKM elements $V_{cs}$ and $V_{cd}$.
  It is well known that the magnitudes of $V_{cs}$ and $V_{cd}$
  are extracted from leptonic and semileptonic charm decays.
  If these nonleptonic decay modes are accurately measured
  in the future, they could offer another determinations or
  constraints of ${\vert}V_{cs}{\vert}$ and ${\vert}V_{cd}{\vert}$.

   \begin{table}[ht]
   \caption{The values of the input parameters, where their
   central values will be regarded as the default inputs
   unless otherwise specified. The numbers in parentheses
   are errors.}
   \label{tab:input-parameter}
   \begin{ruledtabular}
   \begin{tabular}{ccc}
   \multicolumn{3}{c}{mass, width and decay constants
    of the particles \cite{pdg2020} } \\ \hline
    $m_{{\pi}^{0}}$ $=$ $134.98$ MeV,
  & $m_{K^{0}}$ $=$ $497.61$ MeV,
  & $f_{{\pi}}$ $=$ $130.2(1.2)$ MeV, \\
    $m_{{\pi}^{\pm}}$ $=$ $139.57$ MeV,
  & $m_{K^{\pm}}$ $=$ $493.68$ MeV,
  & $f_{K}$ $=$ $155.7(3)$ MeV, \\
    $m_{\eta}$ $=$ $547.86$ MeV,
  & $m_{{\eta}^{\prime}}$ $=$ $957.78$ MeV,
  & $f_{{\eta}_{c}}$ $=$ $398.1(1.0)$ MeV \cite{prd102.054511}, \\
    $m_{c}$ $=$ $1.67(7)$ GeV,
  & $m_{{\eta}_{c}}$ $=$ $2983.9(5)$ MeV,
  & ${\Gamma}_{{\eta}_{c}}$ $=$ $32.1(8)$ MeV, \\ \hline
    \multicolumn{3}{c}{Gegenbauer moments at the scale of ${\mu}$
    $=$ 1 GeV \cite{jhep0605.004}} \\ \hline
    \multicolumn{3}{c}{ $a_{1}^{\pi}$ $=$ $0$, \qquad
    $a_{2}^{\pi}$ $=$ $0.25(15)$, \qquad
    $a_{1}^{K}$ $=$ $0.06(3)$, \qquad
    $a_{2}^{K}$ $=$ $0.25(15)$ }
  \end{tabular}
  \end{ruledtabular}
  \end{table}
   \begin{table}[ht]
   \caption{Branching ratios for the ${\eta}_{c}$
   ${\to}$ $PP^{\prime}$ decays, where the uncertainties
   come from $m_{c}$, ${\mu}_{P}$ and $a_{2}^{P}$,
   respectively.}
   \label{tab:branching-ratio}
   \begin{ruledtabular}
   \begin{tabular}{lc||lc}
     \multicolumn{1}{c}{modes} & branching ratio
   & \multicolumn{1}{c}{modes} & branching ratio \\ \hline
     ${\eta}_{c}$ ${\to}$ $K^{+}K^{-}$
   & $(  1.47^{+  0.14+  0.63+  0.22}_{-  0.13-  0.48-  0.19}){\times}10^{-15}$
   & ${\eta}_{c}$ ${\to}$ $\overline{K}^{0}{\eta}$
   & $(  4.36^{+  0.12+  1.07+ 14.18}_{-  0.12-  0.95-~ 1.60}){\times}10^{-17}$ \\
     ${\eta}_{c}$ ${\to}$ $K^{0}\overline{K}^{0}$
   & $(  1.55^{+  0.15+  0.67+  0.22}_{-  0.13-  0.51-  0.20}){\times}10^{-15}$
   & ${\eta}_{c}$ ${\to}$ $\overline{K}^{0}{\eta}^{\prime}$
   & $(  1.84^{+  0.08+  0.57+  1.89}_{-  0.08-  0.46-  1.24}){\times}10^{-16}$ \\
     ${\eta}_{c}$ ${\to}$ ${\pi}^{+}K^{-}$
   & $(  5.15^{+  0.41+  1.61+ 17.00}_{-  0.38-  1.22-~ 4.32}){\times}10^{-17}$
   & ${\eta}_{c}$ ${\to}$ ${\pi}^{0}{\eta}$
   & $(  9.18^{+  0.85+  4.15+  1.35}_{-  0.77-  3.07-  1.18}){\times}10^{-19}$ \\
     ${\eta}_{c}$ ${\to}$ ${\pi}^{0}\overline{K}^{0}$
   & $(  2.65^{+  0.21+  0.81+  8.55}_{-  0.19-  0.62-  2.22}){\times}10^{-17}$
   & ${\eta}_{c}$ ${\to}$ ${\pi}^{0}{\eta}^{\prime}$
   & $(  5.71^{+  0.53+  2.58+  0.84}_{-  0.48-  1.91-  0.73}){\times}10^{-19}$ \\
     ${\eta}_{c}$ ${\to}$ ${\pi}^{+}{\pi}^{-}$
   & $(  1.38^{+  0.13+  0.62+  0.20}_{-  0.12-  0.46-  0.18}){\times}10^{-18}$
   & ${\eta}_{c}$ ${\to}$ ${\eta}{\eta}$
   & $(  5.08^{+  0.47+  2.18+  1.02}_{-  0.42-  1.54-  0.92}){\times}10^{-16}$ \\
     ${\eta}_{c}$ ${\to}$ ${\pi}^{0}{\pi}^{0}$
   & $(  6.91^{+  0.64+  3.12+  1.02}_{-  0.58-  2.32-  0.89}){\times}10^{-19}$
   & ${\eta}_{c}$ ${\to}$ ${\eta}{\eta}^{\prime}$
   & $(  1.30^{+  0.12+  0.55+  0.26}_{-  0.11-  0.39-  0.24}){\times}10^{-15}$ \\
   & & ${\eta}_{c}$ ${\to}$ ${\eta}^{\prime}{\eta}^{\prime}$
   & $(  9.12^{+  0.84+  3.89+  1.84}_{-  0.76-  2.75-  1.66}){\times}10^{-16}$
  \end{tabular}
  \end{ruledtabular}
  \end{table}

  The branching ratio is defined as follows.
   \begin{equation}
  {\cal B}r \,=\,
   \frac{p_{\rm cm}}{8\,{\pi}\,m_{{\eta}_{c}}^{2}\,
  {\Gamma}_{{\eta}_{c}}}\,
  {\vert} {\cal A}({\eta}_{c}{\to}PP^{\prime}) {\vert}^{2}
   \label{eq:branching-ratio},
   \end{equation}
  where $p_{\rm cm}$ is the center-of-mass momentum of final
  states in the rest frame of the ${\eta}_{c}$ meson.
  The numerical results of branching ratios obtained with
  the input parameters in Table \ref{tab:input-parameter}
  are listed in Table \ref{tab:branching-ratio}.
  Our comments are listed as follows.

  (1)
  Almost all of the decay width of the ${\eta}_{c}$ meson
  should come from the strong interactions.
  The parity violating ${\eta}_{c}$ ${\to}$ $PP^{\prime}$ decays
  can only be induced from the weak interactions.
  For the ${\eta}_{c}$ meson, compared with the strong decays,
  the occurrence probability of the weak decay is very tiny,
  only about $1/{\tau}_{D}\,{\Gamma}_{{\eta}_{c}}$ ${\sim}$
  ${\cal O}(10^{-11})$.
  By analogy with the nonleptonic $B$ meson decays, the pure
  annihilation decay modes are dynamically suppressed by
  helicity. Branching ratios for the pure annihilation
  $B_{s}$ ${\to}$ ${\pi}{\pi}$ decays are about $4$ orders
  of magnitude smaller than those of $B_{s}$ ${\to}$
  $D_{s}{\pi}$ decay \cite{pdg2020}.
  So it is not difficult to imagine that the pure annihilation
  ${\eta}_{c}$ ${\to}$ $PP^{\prime}$ decays should have very
  small branching ratios, about $10^{-15}$ or less.

  (2)
  It is turned out that branching ratios for the ${\eta}_{c}$
  ${\to}$ $PP^{\prime}$ decays within SM are the order of
  $10^{-15}$ ${\sim}$ $10^{-19}$, and far beyond the measurable
  precision limit of BES-III and future STCF.
  However, these branching ratios are not as small as the order
  of $10^{-27}$ expected in Ref. \cite{prd84.032006}.
  More particularly, the branching ratios for the ${\eta}_{c}$
  ${\to}$ $K\overline{K}$ decays can reach up to the order of
  $10^{-15}$ even without a considerable additional contribution
  from new physics (NP) beyond the SM.
  The observation of these decays at any level in the next few
  decades would be a signal of parity violations from new
  sources and a hint of NP.

  (3)
  The ${\eta}_{c}$ ${\to}$ $K\overline{K}$ decays are
  Cabibbo-favored.
  The ${\eta}_{c}$ ${\to}$ ${\pi}\overline{K}$ decays are
  singly Cabibbo-suppressed.
  And the ${\eta}_{c}$ ${\to}$ ${\pi}{\pi}$ decays are
  doubly Cabibbo-suppressed.
  In addition, the decay constants $f_{K}$ $>$ $f_{\pi}$.
  Hence, there is a clear hierarchical pattern among
  branching ratios,
   \begin{equation}
  {\cal B}r({\eta}_{c}{\to}K\overline{K})\, >\,
  {\cal B}r({\eta}_{c}{\to}{\pi}\overline{K})\, >\,
  {\cal B}r({\eta}_{c}{\to}{\pi}{\pi})
   \label{bf-hierarchical-relation}.
   \end{equation}
  If there are enough experimental data to study the
  ${\eta}_{c}$ ${\to}$ $PP^{\prime}$ decays in the future,
  then the Cabibbo-favored ${\eta}_{c}$ ${\to}$ $K\overline{K}$
  decays should have more probabilities to be measured firstly.

  (4)
  The study of the pure annihilation ${\eta}_{c}$ ${\to}$
  $PP^{\prime}$ decays further confirmed that when the two final
  states are the particle and antiparticle pair, such as the
  $K\overline{K}$ and ${\pi}{\pi}$, the factorizable annihilation
  contributions from Fig. \ref{feynman-pqcd} (a) and (b) exactly
  cancel each other because of the isospin symmetry, as analyzed
  in Refs. \cite{npb606.245,prd70.034009,prd85.094003}.
  In addition, the interferences between the nonfactorizable
  annihilation amplitudes for Fig.\ref{feynman-pqcd} (c) and (d)
  are destructive for ${\eta}_{c}$ decays because of the opposite
  signs of the momentum of charm quark propagators.
  The above factors also led to the small branching ratios for the
  ${\eta}_{c}$ ${\to}$ $PP^{\prime}$ decays.

  In summary, the parity violating ${\eta}_{c}$ ${\to}$
  $PP^{\prime}$ decays have been investigated based on the
  available BES-II and BES-III data, while the corresponding
  theoretical study is lack of references for a long time.
  In this paper, considering the experimental needs and the
  high enthusiasms in searching for NP at the intensity frontier,
  the ${\eta}_{c}$ ${\to}$ $PP^{\prime}$ decays are studied
  with the pQCD approach within SM.
  It is found that branching ratios for the concerned processes
  are the order of $10^{-15}$ and less, and beyond the current
  detection capability.
  This study offer a ready reference for future analyses.

  \section*{Acknowledgments}
  The work is supported by the National Natural Science Foundation
  of China (Grant Nos. 11705047, U1632109 and 11547014).

  \begin{appendix}
  \section{Building blocks of decay amplitudes}
  \label{blocks}
  For the sake of convenience in writing, some shorthands are used.
   \begin{equation}
  {\phi}_{{\eta}_{c}}^{a,p}\, =\,
  {\phi}_{{\eta}_{c}}^{a,p}(x_{1})\,e^{-S_{{\eta}_{c}}}
   \label{eq:shorthand-etac},
   \end{equation}
   \begin{equation}
  {\phi}_{P,P^{\prime}}^{a}\, =\,
  {\phi}_{P,P^{\prime}}^{a}(x_{i})\,e^{-S_{P,P^{\prime}}}
   \label{eq:shorthand-p-twist-2},
   \end{equation}
   \begin{equation}
  {\phi}_{P,P^{\prime}}^{p,t}\, =\,
   \frac{{\mu}_{P}}{m_{{\eta}_{c}}}\,
  {\phi}_{P,P^{\prime}}^{p,t}(x_{i})\,e^{-S_{P,P^{\prime}}}
   \label{eq:shorthand-p-twist-3},
   \end{equation}
   \begin{equation}
   C_{i}\, {\cal A}_{jk}(P^{\prime},P) \, =\,
  i\,m_{{\eta}_{c}}^{4}\,
   f_{{\eta}_{c}}\,f_{P}\,f_{P^{\prime}}\,
   \frac{{\pi}\,C_{F}}{N_{c}}\, \big\{
   {\cal A}_{j}(P^{\prime},P,C_{i})
  +{\cal A}_{k}(P^{\prime},P,C_{i}) \big\}
   \label{eq:coefficient-block}.
   \end{equation}
  The subscript $i$ of building block ${\cal A}_{i}$ corresponds
  to the indices of Fig. \ref{feynman-pqcd}.
  The expressions of $\mathcal{A}_{i}$ are written as follows.
   \begin{eqnarray}
  {\cal A}_{a} &=&
  {\int}_{0}^{1}dx_{2}\,dx_{3}
  {\int}_{0}^{\infty}db_{2}\,db_{3}\, {\alpha}_{s}(t_{a})\,
  H_{a}({\alpha}_{g},{\beta}_{a},b_{2},b_{3})\, C_{i}(t_{a})
   \nonumber \\ & & \quad\
   S_{t}(\bar{x}_{2})\, \big\{ {\phi}_{P}^{a}\,
  {\phi}_{P^{\prime}}^{a}\,\bar{x}_{2}
  +2\,{\phi}_{P^{\prime}}^{p}\, \big[
  {\phi}_{P}^{p}\,(1+\bar{x}_{2})+
  {\phi}_{P}^{t}\,x_{2} \big] \big\}
   \label{eq:amp-fig-a},
   \end{eqnarray}
   \begin{eqnarray}
  {\cal A}_{b}  &=& -
  {\int}_{0}^{1}dx_{2}\,dx_{3}
  {\int}_{0}^{\infty}db_{2}\,db_{3}\, {\alpha}_{s}(t_{b})\,
  H_{b}({\alpha}_{g},{\beta}_{b},b_{2},b_{3})\, C_{i}(t_{b})
   \nonumber \\ & & \qquad
   S_{t}(x_{3})\, \big\{ {\phi}_{P}^{a}\,
  {\phi}_{P^{\prime}}^{a}\, x_{3}
   +2\,{\phi}_{P}^{p}\, \big[ {\phi}_{P^{\prime}}^{p}\,(1+x_{3})
   -{\phi}_{P^{\prime}}^{t}\,\bar{x}_{3} \big] \big\}
   \label{eq:amp-fig-b},
   \end{eqnarray}
   \begin{eqnarray}
  {\cal A}_{c}  &=& \frac{1}{N_{c}}\,
  {\int}_{0}^{1}dx_{1}\,dx_{2}\,dx_{3}
  {\int}_{0}^{\infty}db_{1}\,db_{2}\, {\alpha}_{s}(t_{c})\,
  H_{c}({\alpha}_{g},{\beta}_{c},b_{1},b_{2})\,C_{i}(t_{c})
   \nonumber \\ & & \quad\
   \big\{ {\phi}_{{\eta}_{c}}^{a}\, \big[
  {\phi}_{P}^{a}\, {\phi}_{P^{\prime}}^{a}\,(x_{3}-x_{1})
   +\big( {\phi}_{P}^{p}\,{\phi}_{P^{\prime}}^{t} -{\phi}_{P}^{t}\,
  {\phi}_{P^{\prime}}^{p} \big)\,(x_{3}-\bar{x}_{2})
   \nonumber \\ & & \quad \qquad
  +\big( {\phi}_{P}^{p}\, {\phi}_{P^{\prime}}^{p}
  -{\phi}_{P}^{t}\, {\phi}_{P^{\prime}}^{t} \big)\,
   (x_{3}+\bar{x}_{2}- 2\,x_{1}) \big]
   \nonumber \\ & & \quad\
  +{\phi}_{{\eta}_{c}}^{p} \big[ \frac{1}{2}\,
   {\phi}_{P}^{a}\, {\phi}_{P^{\prime}}^{a}\,
   +2\,{\phi}_{P}^{p}\, {\phi}_{P^{\prime}}^{p}
   \big] \big\}_{b_{2}=b_{3}}
   \label{eq:amp-fig-c},
   \end{eqnarray}
   \begin{eqnarray}
  {\cal A}_{d}  &=& \frac{1}{N_{c}}\,
  {\int}_{0}^{1}dx_{1}\,dx_{2}\,dx_{3}
  {\int}_{0}^{\infty}db_{1}\,db_{2}\, {\alpha}_{s}(t_{d})\,
  H_{d}({\alpha}_{g},{\beta}_{d},b_{1},b_{2})\,C_{i}(t_{d})
   \nonumber \\ & & \quad\
   \big\{ {\phi}_{{\eta}_{c}}^{a}\, \big[
  {\phi}_{P}^{a}\, {\phi}_{P^{\prime}}^{a}\,(x_{2}-x_{1})
  +\big( {\phi}_{P}^{p}\,{\phi}_{P^{\prime}}^{t} -{\phi}_{P}^{t}\,
  {\phi}_{P^{\prime}}^{p} \big)\,(x_{3}-\bar{x}_{2})
   \nonumber \\ & & \quad \qquad
  +\big( {\phi}_{P}^{p}\, {\phi}_{P^{\prime}}^{p}
  -{\phi}_{P}^{t}\, {\phi}_{P^{\prime}}^{t} \big)\,
   (2\,\bar{x}_{1}-\bar{x}_{2}-x_{3})\,
   \nonumber \\ & & \quad\
   -{\phi}_{{\eta}_{c}}^{t} \big[ \frac{1}{2}\,
   {\phi}_{P}^{a}\, {\phi}_{P^{\prime}}^{a}\,
   +2\,{\phi}_{P}^{p}\, {\phi}_{P^{\prime}}^{p}
   \big] \big\}_{b_{2}=b_{3}}
   \label{eq:amp-fig-d},
   \end{eqnarray}
   \begin{equation}
   S_{{\eta}_{c}}\, =\, s(x_{1},p_{1}^{+},1/b_{1})
   +2\,{\int}_{1/b_{1}}^{t} \frac{d{\mu}}{{\mu}}{\gamma}_{q}
   \label{eq:sudakov-upsilon},
   \end{equation}
   \begin{equation}
   S_{P}\, =\, s(x_{2},p_{2}^{+},1/b_{2})
              +s(\bar{x}_{2},p_{2}^{+},1/b_{2})
   +2\,{\int}_{1/b_{2}}^{t} \frac{d{\mu}}{{\mu}}{\gamma}_{q}
   \label{eq:sudakov-p},
   \end{equation}
   \begin{equation}
   S_{P^{\prime}}\, =\, s(x_{3},p_{3}^{-},1/b_{3})
                       +s(\bar{x}_{3},p_{3}^{-},1/b_{3})
   +2\,{\int}_{1/b_{3}}^{t} \frac{d{\mu}}{{\mu}}{\gamma}_{q}
   \label{eq:sudakov-p-prime},
   \end{equation}
   \begin{equation}
  {\alpha}_{g}\, =\, m_{{\eta}_{c}}^{2}\,\bar{x}_{2}\,x_{3}
   \label{eq:amp-gluon},
   \end{equation}
   \begin{equation}
  {\beta}_{a}\, =\, m_{{\eta}_{c}}^{2}\,\bar{x}_{2}
   \label{eq:amp-quark-fig-a},
   \end{equation}
   \begin{equation}
  {\beta}_{b}\, =\, m_{{\eta}_{c}}^{2}\,x_{3}
   \label{eq:amp-quark-fig-b},
   \end{equation}
   \begin{equation}
  {\beta}_{c}\, =\, {\alpha}_{g}-
   m_{{\eta}_{c}}^{2}\,x_{1}\,(\bar{x}_{2}+x_{3})
   \label{eq:amp-quark-fig-c},
   \end{equation}
   \begin{equation}
  {\beta}_{d}\, =\, {\alpha}_{g}-
   m_{{\eta}_{c}}^{2}\,\bar{x}_{1}\,(\bar{x}_{2}+x_{3})
   \label{eq:amp-quark-fig-d},
   \end{equation}
  and other definations can be found in Ref. \cite{2101.00549}.
  \end{appendix}

  

\begin{thebibliography}{99}
  \bibitem{pdg2020}
  \href{https://doi.org/10.1093/ptep/ptaa104}
       {P. Zyla {\em et al.} (Particle Data Group),
        Prog. Theor. Exp. Phys. 2020, 083C01 (2020).}
  \bibitem{dataweb}
  \href{http://english.ihep.cas.cn/bes/doc/2250.html}
       {http://english.ihep.cas.cn/bes/doc/2250.html.}
  \bibitem{nimpra614.345}
  \href{https://doi.org/10.1016/j.nima.2009.12.050}
       {M. Ablikim {\em et al.} (BESIII Collaboration),
        Nucl. Instr. Meth. Phys. Res. A 614, 345 (2010).}
  \bibitem{ozi-o}
  \href{https://doi.org/10.1016/S0375-9601(63)92548-9}
       {S. Okubo, Phys. Lett. 5, 165 (1963).}
  \bibitem{ozi-z}
          G. Zweig, CERN-TH-401, 402, 412 (1964).
  \bibitem{ozi-i}
  \href{https://doi.org/10.1143/PTPS.37.21}
       {J. Iizuka, Prog. Theor. Phys. Suppl. 37-38, 21 (1966).}
  \bibitem{prd84.032006}
  \href{https://doi.org/10.1103/PhysRevD.84.032006}
       {M. Ablikim {\em et al.} (BESIII Collaboration),
        Phys. Rev. D 84, 032006 (2011).}
  \bibitem{epjc45.337}
  \href{https://doi.org/10.1140/epjc/s2005-02445-0}
       {M. Ablikim {\em et al.} (BESIII Collaboration),
        Eur. Phys. J. C 45, 337 (2006).}
  \bibitem{rmp68.1125}
  \href{https://doi.org/10.1103/RevModPhys.68.1125}
       {G. Buchalla, A. Buras, M. Lautenbacher,
        Rev. Mod. Phys. 68, 1125, (1996).}
  \bibitem{prl83.1914}
  \href{https://doi.org/10.1103/PhysRevLett.83.1914}
       {M. Beneke, G. Buchalla, M. Neubert, C. Sachrajda,
        Phys. Rev. Lett. 83, 1914 (1999).}
  \bibitem{npb591.313}
  \href{https://doi.org/10.1016/S0550-3213(00)00559-9}
       {M. Beneke, G. Buchalla, M. Neubert, C. Sachrajda,
       Nucl. Phys. B 591, 313 (2000).}
  \bibitem{npb606.245}
  \href{https://doi.org/10.1016/S0550-3213(01)00251-6}
       {M. Beneke, G. Buchalla, M. Neubert, C. Sachrajda,
        Nucl. Phys. B 606, 245 (2001).}
  \bibitem{plb488.46}
  \href{https://doi.org/10.1016/S0370-2693(00)00854-6}
       {D. Du, D. Yang, G. Zhu, Phys. Lett. B 488, 46 (2000).}
  \bibitem{plb509.263}
  \href{https://doi.org/10.1016/S0370-2693(01)00398-7}
       {D. Du, D. Yang, G. Zhu, Phys. Lett. B 509, 263 (2001).}
  \bibitem{prd64.014036}
  \href{https://doi.org/10.1103/PhysRevD.64.014036}
       {D. Du, D. Yang, G. Zhu, Phys. Rev. D 64, 014036 (2001).}
  \bibitem{prl74.4388}
  \href{https://doi.org/10.1103/PhysRevLett.74.4388}
       {H. Li, H. Yu, Phys. Rev. Lett. 74, 4388 (1995).}
  \bibitem{plb348.597}
  \href{https://doi.org/10.1016/0370-2693(95)00174-J}
       {H. Li, Phys. Lett. B 348, 597 (1995).}
  \bibitem{prd52.3958}
  \href{https://doi.org/10.1103/PhysRevD.52.3958}
       {H. Li, Phys. Rev. D 52, 3958 (1995).}
  \bibitem{prd63.074006}
  \href{https://doi.org/10.1103/PhysRevD.63.074006}
       {Y. Keum, H. Li, Phys. Rev. D 63, 074006 (2001).}
  \bibitem{prd63.054008}
  \href{https://doi.org/10.1103/PhysRevD.63.054008}
       {Y. Keum, H. Li, A. Sanda, Phys. Rev. D 63, 054008 (2001).}
  \bibitem{prd63.074009}
  \href{https://doi.org/10.1103/PhysRevD.63.074009}
       {C. L\"{u}, K. Ukai, M. Yang, Phys. Rev. D 63, 074009 (2001).}
  \bibitem{plb555.197}
  \href{https://doi.org/10.1016/S0370-2693(03)00049-2}
       {H. Li, K. Ukai, Phys. Lett. B 555, 197 (2003).}
  \bibitem{epjc73.2437}
  \href{https://doi.org/10.1140/epjc/s10052-013-2437-3}
       {J. Sun, Z. Xiong, Y. Yang, G. Lu,
        Eur. Phys. J. C 73, 2437 (2013).}
  \bibitem{2012.10581}
  \href{http://arxiv.org/abs/2012.10581}
       {Y. Yang, L. Lang, X. Zhao, J. Huang, J. Sun, arXiv:2012.10581}
  \bibitem{jhep0605.004}
  \href{https://doi.org/10.1088/1126-6708/2006/05/004}
        {P. Ball, V. Braun, A. Lenz, JHEP 0605, 004 (2006).}
  \bibitem{prd65.014007}
  \href{https://doi.org/10.1103/PhysRevD.65.014007}
       {T. Kurimoto, H. Li, A. Sanda, Phys. Rev. D 65, 014007 (2001).}
  \bibitem{prd65.074001}
  \href{https://doi.org/10.1103/PhysRevD.65.074001}
       {D. Du, H. Gong, J. Sun, D. Yang, G. Zhu,
        Phys. Rev. D 65, 074001 (2002).}
  \bibitem{prd65.094025}
  \href{https://doi.org/10.1103/PhysRevD.65.094025}
       {D. Du, H. Gong, J. Sun, D. Yang, G. Zhu,
        Phys. Rev. D 65, 094025 (2002).}
  \href{https://doi.org/10.1103/PhysRevD.66.079904}
       {Erratum, Phys. Rev. D 66, 079904 (2002).}
  \bibitem{prd68.054003}
  \href{https://doi.org/10.1103/PhysRevD.68.054003}
       {J. Sun, G. Zhu, D. Du,
        Phys. Rev. D 68, 054003 (2003).}
  \bibitem{npb675.333}
  \href{https://doi.org/10.1016/j.nuclphysb.2003.09.026}
       {M. Beneke, M. Neubert, Nucl. Phys. B 675, 333 (2003).}
  \bibitem{npb774.64}
  \href{https://doi.org/10.1016/j.nuclphysb.2007.03.020}
       {M. Beneke, J. Rohrer, D. Yang, Nucl. Phys. B 774, 64 (2007).}
  \bibitem{prd90.054019}
  \href{https://doi.org/10.1103/PhysRevD.90.054019}
       {Q. Chang, J. Sun, Y. Yang, X. Li, Phys. Rev. D 90, 054019 (2014).}
  \bibitem{prd91.074026}
  \href{https://doi.org/10.1103/PhysRevD.91.074026}
       {Q. Chang, X. Hu, J. Sun, Y. Yang, Phys. Rev. D 91, 074026 (2015).}
  \bibitem{plb740.56}
  \href{http://dx.doi.org/10.1016/j.physletb.2014.11.027}
       {Q. Chang, J. Sun, Y. Yang, X. Li, Phys. Lett. B 740, 56 (2015).}
  \bibitem{plb743.444}
  \href{http://dx.doi.org/10.1016/j.physletb.2015.03.001}
       {J. Sun, Q. Chang, X. Hu, Y. Yang, Phys. Lett. B 743, 444 (2015).}
  \bibitem{plb504.6}
  \href{https://doi.org/10.1016/S0370-2693(01)00247-7}
       {Y. Keum, H. Li, A. Sanda, Phys. Lett. B 504, 6 (2001).}
  \bibitem{prd70.034009}
  \href{https://doi.org/10.1103/PhysRevD.70.034009}
       {Y. Li, C. L\"{u}, Z. Xiao, X. Yu, Phys. Rev. D 70, 034009 (2004).}
  \bibitem{prd85.094003}
  \href{https://doi.org/10.1103/PhysRevD.85.094003}
       {Z. Xiao, W. Wang, Y. Fan, Phys. Rev. D 85, 094003 (2012).}
  \bibitem{prd76.074018}
  \href{https://doi.org/10.1103/PhysRevD.76.074018}
       {A. Ali, G. Kramer, Y. Li {\em et al.}, Phys. Rev. D 76, 074018 (2007).}
  \bibitem{prd88.014043}
  \href{https://doi.org/10.1103/PhysRevD.88.014043}
       {K. Wang, G. Zhu, Phys. Rev. D 88, 014043 (2013).}
  \bibitem{prd58.114006}
  \href{https://doi.org/10.1103/PhysRevD.58.114006}
       {Th. Feldmann, P. Kroll, B. Stech, Phys. Rev. D 58, 114006 (1998).}
  \bibitem{prd89.114019}
  \href{https://doi.org/10.1103/PhysRevD.89.114019}
       {J. Sun, Y. Yang, Q. Chang, G. Lu, Phys. Rev. D 89, 114019 (2014).}
  \bibitem{prd102.054511}
  \href{https://doi.org/10.1103/PhysRevD.102.054511}
       {D. Hatton, C. Davies, B. Galloway {\em et al.},
        Phys. Rev. D 102, 054511 (2020).}
  \bibitem{2101.00549}
  \href{http://arxiv.org/abs/2101.00549}
       {Y. Yang, M. Duan, J. Lu, J. Huang, J. Sun, arXiv:2101.00549}
  \end{thebibliography}
  \end{document}